# How to Investigate the Historical Roots and Evolution of Research Fields in China? A Case Study on iMetrics Using RootCite

Xin Li † (0000-0002-8169-6059)
School of Information Management, Wuhan University, Wuhan, Hubei, China
School of Information, The University of Texas at Austin, Austin 78701, TX, U.S.A.

Qiang Yao † (0000-0002-4408-087X)
School of Political Science and Public Administration, Wuhan University, Wuhan 430074, China

Xuli Tang * (0000-0002-1656-3014)
School of Information Management, Wuhan University, Wuhan 430074, China
School of Informatics, Computing, and Engineering, Indiana University, Bloomington, IN, U.S.A.

Qian Li
Department of Information Management, Peking University, Beijing 100871, China

Mengjia Wu
Centre for Artificial Intelligence, Faculty of Engineering and Information Technology, University of Technology Sydney, Ultimo 2007, NSW, Australia

**† *Xin Li and Qiang Yao contributed equally to this work.***

**\* *Corresponding author: Xuli Tang, e-mail: xulitang@whu.edu.cn***

**Abstract**

This paper aimed to provide an approach to investigate the historical roots and evolution of research fields in China by extending the reference publication year spectroscopy (RPYS). RootCite, an open source software accepts raw data from both the Web of Science and the China Social Science Citation Index (CSSCI), was developed using python. We took iMetrics in China as the research case. 5,141 Chinese iMetrics related publications with 73,376 non-distinct cited references (CR) collected from the CSSCI were analyzed using RootCite. The results showed that the first CR in the field can be dated back to 1882 and written in English; but the majority (64.2%) of the CR in the field were Chinese publications. 17 peaks referring to 18 seminal works (13 in English and 5 in Chinese) were located during the period from 1900 to 2017. The field shared the same roots with that in the English world (e.g., Lotka's law and Garfield's "Citation Indexes") but has its own characteristics, and it was then shaped by contributions from both the English world (e.g., Small's "Co-citation" and Callon et al.'s "Co-word analysis" ) and China (e.g., Qiu's "Bibliometrics" and Su's "CSSCI"). The five Chinese works have played irreplaceable and positive roles in the historical evolutionary path of the field, which should not be ignored, especially for the evolution of the field. This research demonstrated how RootCite aided the task of identifying the origin and evolution of research fields in China, which could be valuable for extending RPYS for countries with other languages.

**Keywords** iMetrics in China; Reference Publication Year Spectroscopy; RPYS; RootCite; Algorithm historiography; China Social Science Index (CSSCI)

## Article highlights

1. This paper introduced RootCite (a Python-based tool for RPYS analysis on Chinese publications), showcased and verified how it can aid the task of locating seminal works in the historical evolutionary path of a research field in China.

2. This paper examined the historical roots and seminal works of iMetrics in China using RootCite, which could be a valuable example for extending RPYS for countries with other languages.

3. A total of 17 significant peaks referring to 18 seminal works (13 in English and 5 in Chinese) were identified during 1900-2017, which is characterized by three stages: budding (before 1970), formation (1971-2000), and development and expansion (2001-2017). IMetrics in China rooted in the same contributions as the English world but it has its own characteristics. The pioneers of iMetrics in China paid more attetion on applied aspect (e.g, paper networks and citation analysis), while the English world have deeped into the basic theory of this field. Several Chinese works (e.g. Qiu's "Bibliometrics" and Su's CSSCI) have an irreplaceable and positive effects on the development and evolution of iMetrics in China.

## Introduction

Algorithmic historiography (AH), originally proposed by Garfield (1964), has recently been intensively researched with the explosive growth in the number of research articles (Linnenluecke and Griffiths 2013; Olijnyk 2015; Elango et al. 2016). One of the most promising approaches in AH is Reference Publication Year Spectroscopy (RPYS), introduced by (Leydesdorff et al. 2014; Marx et al. 2014; Marx and Bornmann 2014). From the perspective of cited references, RPYS can be employed to identify seminal papers which are the most frequently cited in a specific reference publication year, even the early works that published earlier than the existence of the field. Meanwhile, different from the traditional qualitative (e.g., system review) or quantitative approaches (e.g., citation counts), it can locate the seminal articles for a research field in a more objective way by considering opinions from all scientists (via all the reference cited in their papers) in the field.

    The previous studies relating to RPYS mainly focus on two aspects, one of which is the application of RPYS to locate the seminal papers for a research field, a research topic or a scientist. Up to now, it has been successfully employed to several research fields or topics, such as global positioning system (Comins and Hussey 2015), depression (Geraei et al. 2018), health equity (Yao et al. 2019), iMetrics (Leydesdorff et al. 2014; Li 2019), tribology (Elango et al. 2016), and the Darwin finches (Marx and Bornmann 2014). RPYS was also used to identify the research fronts and sleeping beauties by (Comins and Leydesdorff 2016), and seminal works of a scientist, such as Eugene Garfield (Bornmann, Haunschild and Leydesdorff 2018) and Judit Bar-llan (Bornmann and Leydesdorff 2020).

Another important research direction of RPYS is the development and optimization of the tools for RPYS analysis. The first tool available for RPYS analysis was the RPYS.exe developed by (Marx et al. 2014), which can only compute the standard RPYS without the visualization function. Then, (Thor et al. 2016) introduced the first version of a user-friendly program CRExplorer with powerful graph making for standard RPYS. In the same year, (Comins and Leydesdorff 2016) proposed Multi-RPYS and designed a web-based tool called RPYS i/o that can compute and visualize standard RPYS and Multi-RPYS, but it didn't accept data exceeding 15MB. In most recently, (McLevey and McIlroy-Young 2017) introduced a full-featured python package named metaknowledge that accepts a large scale of data from academic databases. Moreover, a web-based tool called Patent citation spectroscopy (PCS) available for identifying seminal patents were introduced by (Comins et al. 2018). Besides, (Thor et al. 2018) further extended the CRExplorer with a new feature, that is, the sequence of citation counts of a cited reference over the citing years, which can be used for identifying the "hot papers" and the "sleeping beauties" in a research field.

However, most of these studies or tools tried to identity the historical roots of a research field based on the datasets collected only from the Web of Science (WoS), in which the publications written in English; few studies have paid attention to the historical roots of a research field in a country with other languages, such as China. Meanwhile, most bibliometricians outside China have limited knowledge of Chinese science and technology system and lack a comprehensive understanding of scientometric resources and methods for analyzing Chinese science (Waltman et al. 2019). Although China has currently become one of the world-leading scientific nations with enlightened research policies and sufficient funding, many of the research fields in China was established from scratch (Hu 2019). Hence, it becomes a significant research issue to investigate the historical roots and evolution of research fields in China.

The aim of this paper is to extend the RPYS method to locate the seminal works in the historical evolutionary path of a research field in China. We used iMetrics in China as a case study and employed the China Social Science Citation Index (CSSCI) (Su et al. 2014) as our data source. IMetrics is one of the most significant branches in library and information science (LIS), and has been defined as a research field with similar purpose and methods, including bibliometrics, informetrics, scientometrics and webometrics (Milojević and Leydesdorff 2013). It was the year 1964 when the term "bibliometrics" was first introduced into china; and there were few studies on iMetrics published in Chinese before 1970, which was considered as the budding period of iMetrics in China (Lamirel et al. 2020; Qiu et al. 2003). In 1983, researchers in Wuhan University offered the first course of "bibliometrics" for undergraduates in China, and they also wrote the first textbook of bibliometrics in Chinese (Lamirel et al. 2020; Qiu et al. 2003). Since then, iMetrics research has started to grow rapidly in China. In 2017, the 16[th] International Conference on Scientometrics and Informetrics (ISSI) jointly organized by Wuhan University was held in Wuhan China, indicating that China has become one of the research centers of iMetrics in the world (Yang et al. 2019). Although iMetrics was originally introduced into China in the early1970s, however, the historical roots could have emerged much earlier. Therefore, it is a suitable subject for us to validate the extended RPYS method on a Chinese data source. In addition, the historical roots of iMetrics in the English world have been researched by (Leydesdorff et al. 2014) with RPYS.exe, using the papers collected from the journals including Scientometrics, Informetrics and JASIST, which provided us with opportunity to compare the differences between the historical roots and evolution of iMetrics in China and the English world.

## Data and Methodology

### Data collection

The China Social Science Citation Index (CSSCI) is one of the most authority academic databases with citation indexes in China that contains the most influential Chinese journals in humanities and social sciences (Su et al. 2014). Therefore, the dataset used in this study were collected from the online version of CSSCI operated by the Institute for Chinese Social Sciences Research and Assessment, Nanjing, China, on June 20, 2020.

The search strategy was based on a set of search terms, which were selected by using a sophisticated approach called as "interactive query formulation" (Wacholder 2011). Specifically, we first established a collection of 2376 core articles on iMetrics by searching the articles whose titles contain "文献计量"(bibliometrics), or "信息计量" (informetrics), or "科学计量"(scientometrics), or "网络计量"(webometrics) or their synonyms. Then, we analyzed the words of titles, abstracts, and author-selected keywords of the core articles, to obtain a list of the most relevant and frequent domain-specific vocabularies of iMetrics in China. Thereafter, based on the vocabularies and suggestions from domain experts, we extended the former search query as follows (translated in English and the Chinese version can be seen in the Supplementary Information S1).

*All Fields = ('bibliometrics' OR 'webometrics' OR 'informetrics' OR 'scientometrics' OR 'knowledge metrics' OR 'citation analysis' OR 'altmetrics' OR 'co-word analysis' OR 'journal evaluation' OR 'paper evaluation' OR 'scientific evaluation' OR 'academic impact' OR 'h index' OR 'university rank' OR 'open access') AND publication year = (1998-2017) AND article type = ('article' or 'review').*

A total of 5,216 articles were retrieved from the China Social Science Citation Index (CSSCI) covering the period of 1988-2017. Considering that "academic impact" and "university rank" could also be used in other fields (such as education), we manually checked the 337 articles that were related to the two search terms and excluded 75 articles that were irrelevant to iMetrics. Finally, 5141 iMetrics related articles were downloaded from the CSSCI in the plain text format. An example of the bibliographic information of an article downloaded from the CSSCI can be found in the Supplementary Information S2. A text parser was then developed using python language and the spaCy (https://spacy.io/) to extract meta data of these articles, such as titles, publication year, cited references, authors and abstracts. There were total 73,376 non-distinct cited references of these articles, in which the number of Chinese cited references was 47,126 (accounting for 64.2%); however, the earliest cited reference was published in 1882 and written in English. The detailed descriptive information of our dataset is shown in Table 1.

Table 1. The descriptive information of our dataset.

| *Item* | *Value* |
|---|---|
| Number of publications | 5,141 |
| Publication Year | 1998-2017 |
| Average Number of References | 14.3 |
| Total Number of References | 73,376 |
| Reference Publication Year | 1822-2017 |
| Number of References in Chinese | 47,126 |
| Number of References in English | 26,250 |

**Method and toolkit**

Reference publication year spectroscopy (RPYS) was originally proposed by (Marx et al. 2014) for tracking the historical roots and seminal works of a specific domain, researcher or topic. Compared to traditional methods, such as the simple citation counts or the HistCite (Garfield 2004), RPYS takes the negative effects of citations from other domains as well as the publication time into consideration, using the visual peaks and 5-year deviations from the perspective from cited references (Marx and Bornmann 2014). The general workflow entails data collection and preprocessing, standard RPYS curve plotting, important RPYs identification, and seminal works identification for a specific RPY (Wray and Bornmann 2015).

In previous studies relating to RPYS, tools including RPYS.exe (Marx et al. 2014), RPYS i/o (Comins and Leydesdorff 2016), CitedReferencesExplorer (CRExplorer) (Thor et al. 2016), and Metaknowledge (McLevey and McIlroy-Young 2017) were successively developed for conducting RPYS analysis on the datasets in the form of plain text downloaded from the WoS; however, none of these tools can be applied to publications written in Chinese. In this study, we extended the RPYS.exe and developed a tool called RootCite using python, which accepts publications from both the WoS and the CSSCI, to investigate the historical roots and evolution of iMetrics in China.

The graphic user interface of RootCite and the main steps for how to conduct RPYS analysis with it are shown in the Figure 1 and 2, respectively. The detail information about the procedures for RootCite is provided in the Supplementary Information S3. The source code of RootCite can be freely download from the link (https://github.com/isxinli/RootCite). Using the CSSCI data (or WoS data) as the input, RootCite returns two files, that is, "median_cssci.csv" ("median_wos.csv") and "result_cssci.csv" ("result_wos.csv"). The "median_cssci.csv" contains the number of cited references according to the reference publication year (RPY), as well as the differences from 5-year (i.e., the current, the two previous and the two following years) median per RPY. This file can be opened by Excel for plotting the standard RPYS for identifying the important RPYs for a specific domain. Then, the "result_cssci.cssv", in which the number of cited times and the details for a specific cited reference are provided, can be used for identifying seminal works in an important RPY. It should be noted that the results of RPYS analysis must be explained in the historical context and validated by domain experts (Leydesdorff et al. 2014; Marx et al. 2014; Marx and Bornmann 2014).

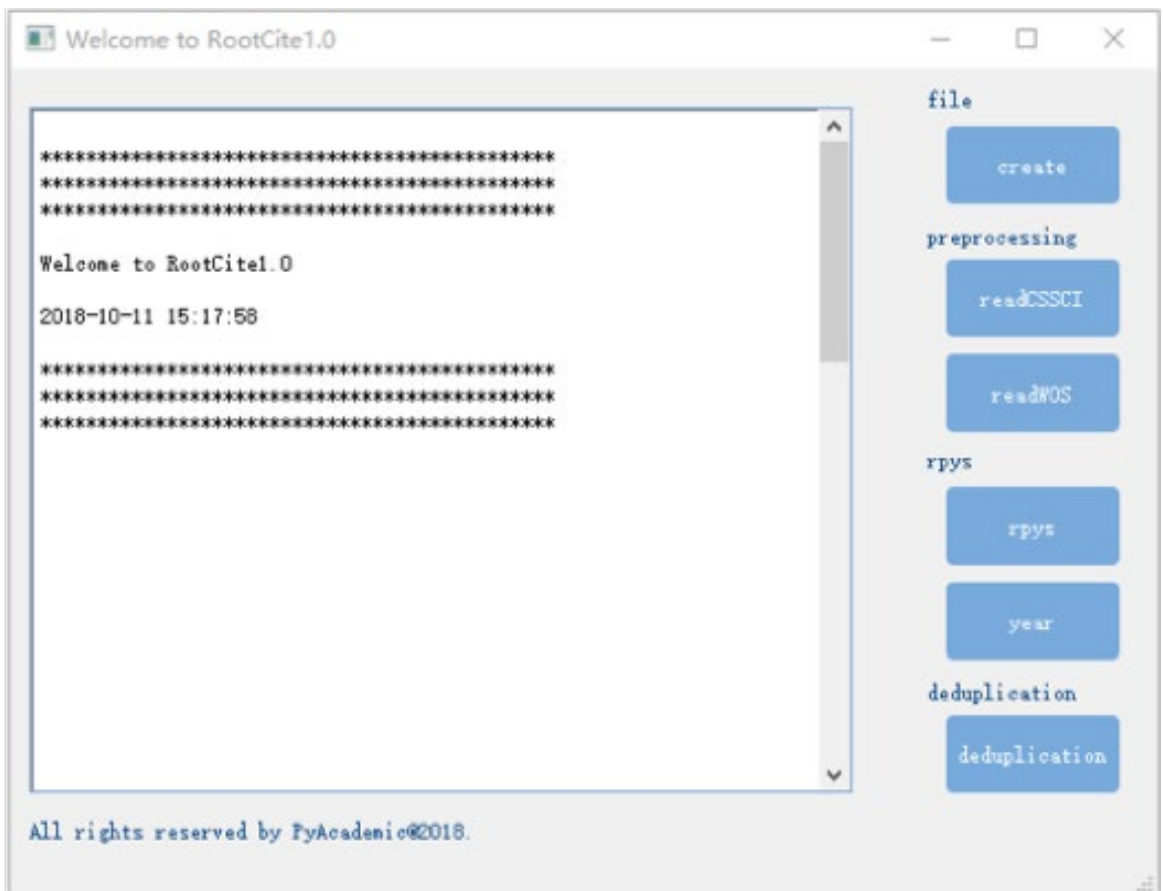

Figure 1. The simple graphical user interface of RootCite.

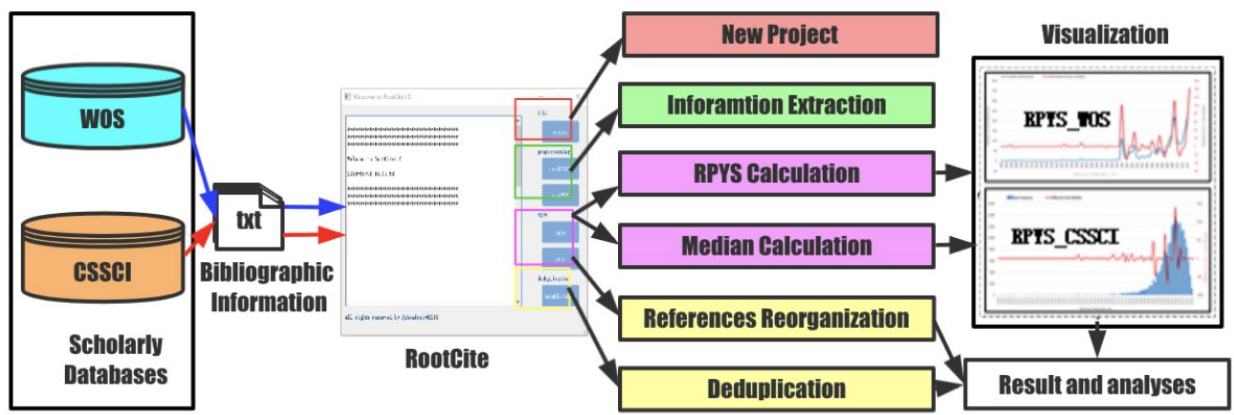

Figure 2 The main steps for investigating the historical roots and evolution of a scientific domain with RootCite.

It is worth noting that a cited reference could have multiple variants in the CSSCI database because of the different reference formats requested by different journals. To normalize the cited references, in the RPYS.exe and the CRExplorer, (Leydesdorff et al. 2014; Marx et al. 2014) used the Levenshtein algorithm (i.e., edit distance) to measure the text similarity between cited reference strings. They matched and clustered two reference strings when the similarity of them reached the threshold of 0.75. This process is time-consuming and memory consuming by systematically compared between thousands or even millions of strings (McLevey and McIlroy 2017). McLevey and McIlroy (2017) created an identification strings for each cited reference by using its author, year and journal, significantly reducing the runtime from hours to minutes for large datasets. However, the accuracy of this method is not high since it is possible that different cited references have the same identification string. For example, the same authors could have published more than one article in the same journal in the same year. Therefore, in the "deduplication" module of RootCite (Figure 1 and the Supplementary Information S3), we employed the Minhash algorithm, an efficient algorithm of text similarity calculating with lower time complexity and space complexity (Chum et al. 2008), to match and cluster cited references when their similarity reached the threshold of 0.85.

## Results and analyses

### Overview of iMetrics in China

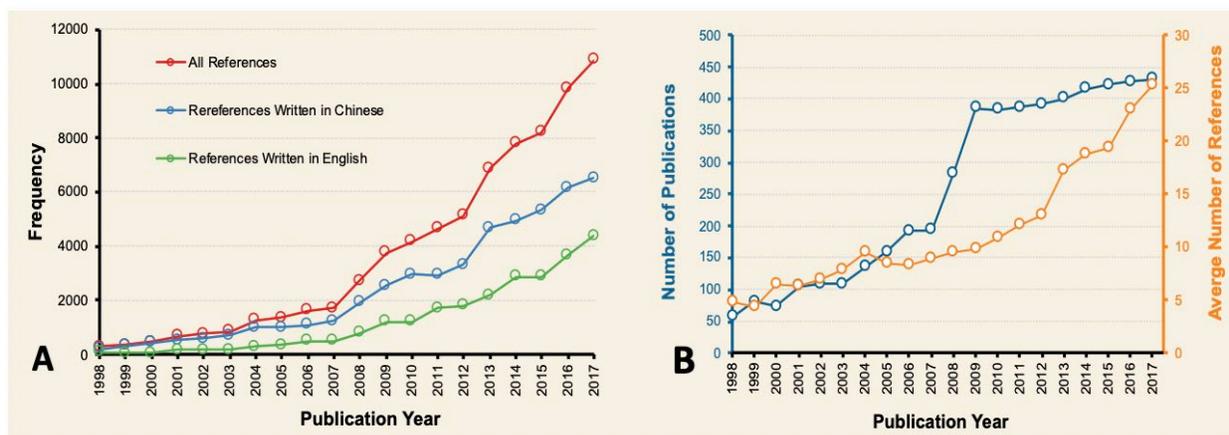

Figure 3. The annual distribution of the number of cited references for iMetrics in China according to the publication year: A. the annual frequency of references in two languages for iMetrics in China (1998-2017); B. the average number of references per paper for iMetrics in China (1998-2017).

The annual distribution for the number of cited references for iMetrics in China during the period of 1998-2017 is shown in the Figure 3. The total number of references cited in iMetrics dramatically increased from 274 in 1998 to more than 10,000 in 2017, more than 30 times (Figure 3A). The blue curve denotes the number of cited references written in Chinese as a function of the publication year, from which we observe that, as time goes by, this number grew steadily, especially after the year 2007. The green line representing the number of cited references written in English, significantly increased from less than 80 in 1998 to 4368 in 2017, more than 50 times. Meanwhile, the blue line has been always above the green one overall, illustrating that the number of the cited references written in Chinese have been more than that in English. Moreover, the gap between the number of references in two languages have shown a clear increasing trend, from 122 in 1988 to 2147 in 2017.

Figure 3B represents the average number of cited references and the number of publications according to the publication year (1998-2017), from which we find that, with the growth of the number of publications for iMetrics in China, the average number of cited references also increased steadily, from 4.94 in 1998 to 25.3 in 2017. We also observe that less than 100 publications related to iMetrics were yearly produced before the year 2000, indicating the formation stage of the iMetrics research in China (Qiu et al. 2003). In 2001-2009, the number of publications had grown dramatically, which indicates the high-speed development stage of the iMetrics in China. Thereafter, the annual number of publications started to keep steadily with small fluctuations, which means iMetrics in China entered into its maturity stage.

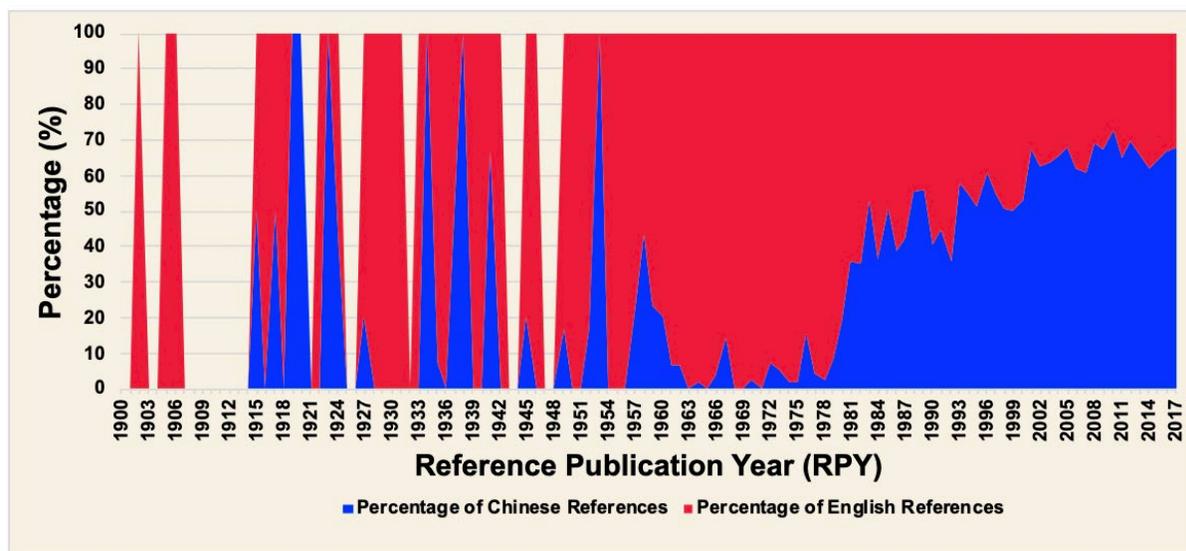

Figure 4. The percentage of references in two languages of iMetrics in China according to the reference publication year (1900-2017).

Figure 4 shows the distribution of cited references of the iMetrics in China with a 100% stacked area graph, in which the cited references written in two different languages respectively as a percentage of all cited references of the iMetrics in China is presented. Before 1960, the percentage of cited references in two languages fluctuated dramatically and no obviously trends can be observed, since the number of publications and their references was very small and unstable (Qiu et al. 2003). From 1961 to 1982, the red area denoting the percentage of cited references written in English was far more than that of cited references written in Chinese represented by the blue area, which indicates that at the early stage of iMetrics research, pioneers in China tended to absorb the advanced experience from abroad. At 1983, the percentage of cited references written in Chinese firstly exceeded 50%, which reflects the development of the iMetrics research in China. After 2000, the percentages of Chinese references have been above 60%, indicating the Chinese iMetrics research experienced its high-speed development stage and entered into its maturity.

Based on the related studies (Qiu et al. 2003; Yang et al. 2019) and the above observation, we can conclude that, for identifying the historical roots and evolution of iMetrics in China, the contribution of China should not be ignored, and it is necessary to take the Chinese dataset into consideration.

**The Reference Publication Year Spectroscopy of iMetrics in China**

We employed RootCite to conduct RPYS analysis on the dataset of iMterics in China from the CSSCI. The standard RPYS graph of iMerics in China during the period of 1900-2017 is presented in Figure 5, in which the red and blue lines denote the number of cited references and the differences from 5-year median (including the first two years, the current year and the next two years), respectively. We observe that the RPY 2008 obtained the most cited times (4738), indicating the intensive relevant contributions to the iMetrics research in China. The positive peaks of the blue curve densely distributed between the RPYs 1995-2008. Nevertheless, some earlier RPYs appears to be significant too, for examples, 1926, 1934 or 1988. We divided the RPYs into three periods, i.e., before 1970, 1971-2000 and 2000-2017, to identify the seminal works in the historical evolutionary path of iMetrics research in China. There are two reasons for that. First, according to the previous related works (Lamirel et al. 2020; Qiu et al. 2003), iMetrics was originally introduced into China in the early 1970s; however, the historical roots could have emerged much earlier. Hence, we identified the historical roots of iMetrics in China before 1970. Second, our analysis on the yearly number of iMetrics related articles in China (Figure 3B) indicated that, the period of 1971-2000 was the formation stage of the iMetrics research in China, with few articles yearly published (Qiu et al. 2003); and then, in the period of 2001-2017, iMetrics in China had gone through a high-speed development and entered into its maturity (Lamirel et al. 2020; Yang et al. 2019). Therefore, we identified the seminal works in the evolution of iMetrics in China during the two periods (1971-2000 and 2001-2017).

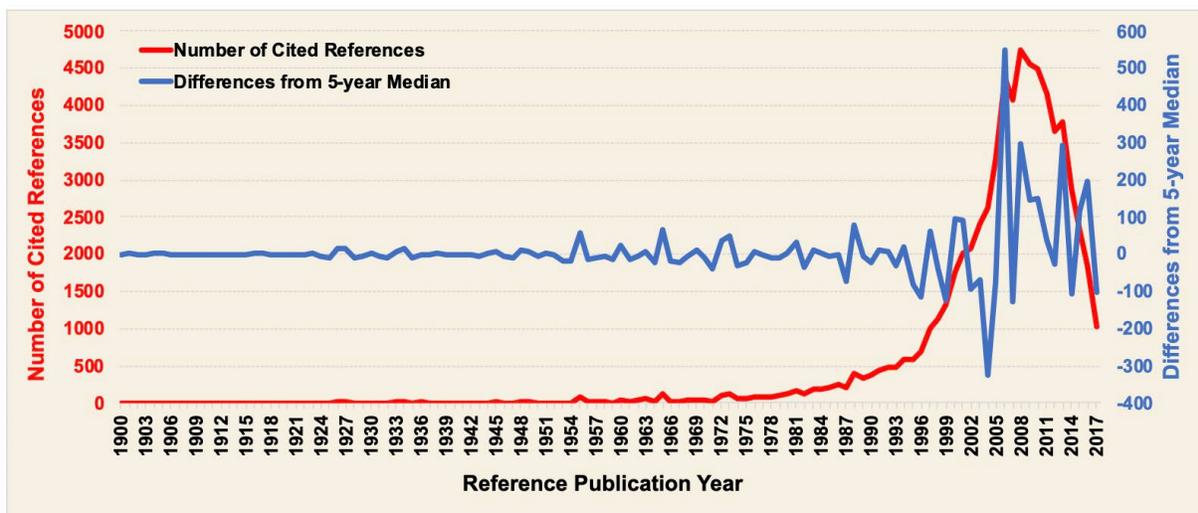

Figure 5. The reference publication year spectroscopy of iMetrics in China (1900-2017).

*Stage one (before 1970): the budding of iMetrics in China*

As shown in Figure 6, we can easily find that there are five major peaks (1926, 1934, 1955, 1960 and 1965) during the seven decades (1900-1970). And if we conduct a more careful analysis of the detail information in the file median_cssci.csv and result_cssci.csv, other three significant peaks can also be found, including 1917, 1944 and 1963. The details about all the eight peaks are presented in Table 2, from which we see that all these seminal cited references were published on journals and written in English. The first peak refers to a paper published on Science Progress, in which Cole J and Eales B conducted a statistical analysis of anatomy papers (Cole and Eales 1917). This work accounts for the 100% citation rate and is generally considered as the first bibliometric study in the world.

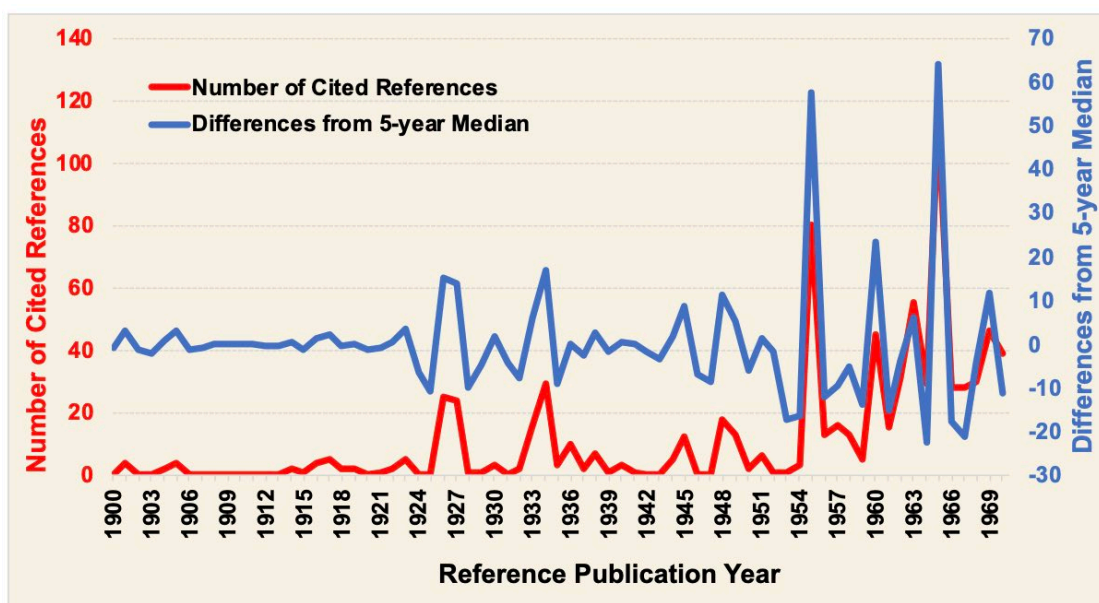

Figure 6. The reference publication year spectroscopy of iMetrics in China (1900-1970). Notes: the seminal works related to iMetrics in China published in 1917, 1926, 1934, 1944, 1955, 1960, 1963 and 1965.

The second peak (1926) and the third peak (1934) respectively first proposed Lotka's law and Bradford's law, which are known as the two most basic laws in the field of bibliometrics. Lotka (1926), accounting for the total citation rate of 75%, firstly uncovered the relationship between authors and the number of their publications; while Bradford (1934) that were cited 27 times, are widely utilized for identifying core journals in a scientific domain. We also noted that, similar with the result of (Hou 2017), the Zipf's law (i.e., the law of word frequency distribution) that is also known as one of the most basic laws in Bibliometrics was not recognized in this study. By rechecking our dataset, we found that this may be because most of the Chinese articles mentioned the Zipf's law had cited the two Chinese books 《文献计量学》 ("Bibliometrics") (Qiu 1988) and 《信息计量学》 ("Informetrics") (Qiu 2007) instead of the classical work published by G.K. Zipf in the 1940s (Zipf 1949). Besides, this finding also indicates that studies on iMetrics in China may have focused on the article level more than the word level (Qiu 2003), especially in its budding period.

Table 2. Details about the significant peaks for iMetrics in China before 1970.

| NO | RPY | Most Cited Reference | Percentage of citations (%) | Document Type |
|---|---|---|---|---|
| 1 | 1917 | Cole, F. J., & Eales, N. B. (1917). The history of comparative anatomy: Part I.—A statistical analysis of the literature. Science Progress (1916-1919), 11(44), 578-596 | 100.0 | Journal |
| 2 | 1926 | Lotka, A. J. (1926). The frequency distribution of scientific productivity. Journal of the Washington academy of sciences, 16(12), 317-323. | 75.0 | Journal |
| 3 | 1934 | Bradford, S. C. (1934). Sources of information on specific subjects. Engineering, 137, 85-86. | 86.2 | Journal |
| 4 | 1944 | Gosnell, C. F. (1944). Obsolescence of books in college libraries. college & research library. | 80.0 | Journal |
| 5 | 1955 | Garfield, E. (1955). Citation indexes for science. Science, 122, 108-111. | 98.6 | Journal |
| 6 | 1960 | Burton, R. E., & Kebler, R. W. (1960). The "half-life" of some scientific and technical literatures. American documentation, 11(1), 18-22. | 41.9 | Journal |
| 7 | 1963 | Kessler, M. M. (1963). Bibliographic coupling between scientific papers. American documentation, 14(1), 10-25. | 48.2 | Journal |
| 8 | 1965 | Price, D. J. D. S. (1965). Networks of scientific papers. Science, 510-515. | 69.7 | Journal |

The fourth peak refers to 1944 with a paper on literature obsolescence written by Gosnell, in which the phenomenon of the reduction in the value of scientific literature overtime was originally investigated (Gosnell 1944). After six years, a measurement of literature obsolescence called "half-life" drawing on a concept from the domain nuclear physics was presented by Burton and Kebler (1960), referring to the sixth peak and accounting for 41.9% of the total citations.

The fifth peak happened in the RPY 1955 in which Garfield E who was famous for the father of scientometrics published an article entitled "Citation indexes for science" on the most influential journal Science (Garfield E 1955). This work is widely considered as the initiation of the method of citation analysis and is the foundation of the Science Citation Index (SCI), which is an important database for the iMetrics research in the world.

The seventh peak is in the RPY 1963 that is because of the creation of bibliographic coupling and its application for measuring the static correlation between two scientific papers. The more bibliographic couples exist, the more relevant the two papers are (Kessler 1963). The last peak happened in the year 1965, one of the two most outstanding peaks (1955 and 1965) during 1900-1971, in which Price D published an article entitled "Networks of scientific paper" on Science. In his paper, Price D pointed out that the patterns of biographic information could be utilized for detecting the essence of the scientific research front (Price 1965).

### *Stage two (1971-2000): the formation of iMetrics in China*

Figure 7 shows the reference publication year spectroscopy of iMetrics in China in the period of 1971-2000. There results illustrate that there are six significant peaks for the iMetrics in China during this period. The details about these peaks are shown in Table 3, from which we see that two peaks refer to cited references written in Chinese and Qiu J authored the both, while the remaining peaks are all in English.

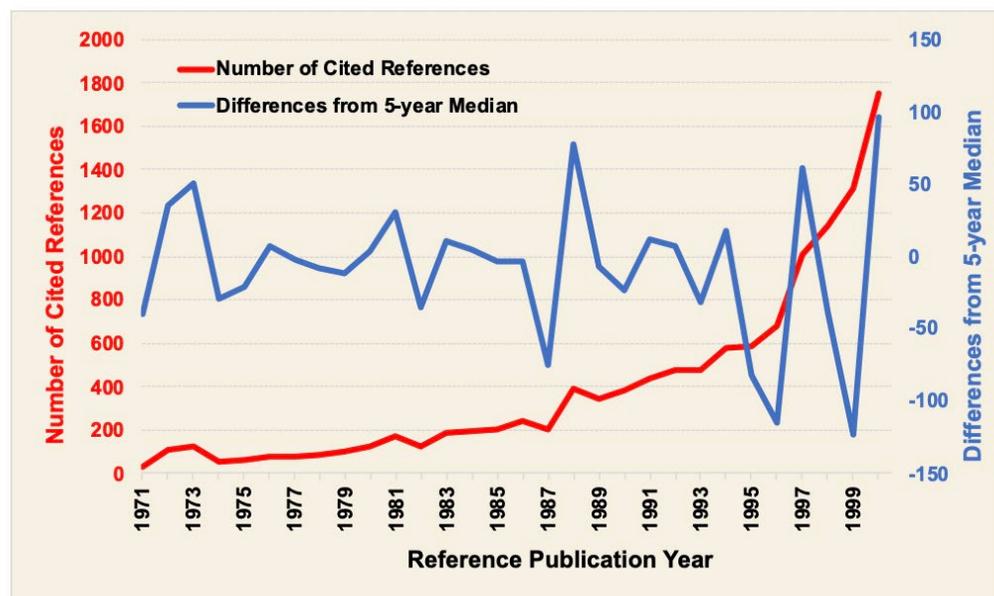

Figure 7. The reference publication year spectroscopy of iMetrics in China (1971-2000). Notes: the seminal works related to iMetrics in China published in 1973, 1981, 1983, 1988, 1997 and 2000.

Table 3. Details about the significant peaks for iMetrics in China between 1971-2000.

| NO | RPY | Most Cited Reference | Percentage of citations (%) | Document Type |
|---|---|---|---|---|
| 1 | 1973 | Small, H. (1973). Co-citation in the scientific literature: A new measure of the relationship between two documents. Journal of the American Society for information Science, 24(4), 265-269. | 56.0 | Journal |
| 2 | 1981 | White, H. D., and Griffith, B. C. (1981). Author cocitation: A literature measure of intellectual structure. Journal of the American Society for information Science, 32(3): 163-171 | 28.1 | Journal |
| 3 | 1983 | Callon, M., Courtial, J. P., Turner, W. A., & Bauin, S. (1983). From translations to problematic networks: An introduction to co-word analysis. Information (International Social Science Council), 22(2), 191-235. | 47.3 | Journal |
| 4 | 1988 | 邱均平,文献计量学,北京:科学技术文献出版社,1988. *[Qiu J. (1988). Bibliometrics. Science and Technology Literature Publishing House, Beijing China.]* | 47.4 | Book |
| 5 | 1997 | Almind, T. C., & Ingwersen, P. (1997). Informetric analyses on the world wide web: methodological approaches to 'webometrics'. Journal of documentation, 53(4), 404-426. | 23.4 | Journal |
| 6 | 2000 | 邱均平,信息计量学系列论文 (一至六), 情报理论与实践, 2000. *[Qiu J. (2000). "Informetrics" (1-6). Information Studies: Theory & Application.]* | 35.6 | Journal |

The first peak in the period of 1971-2000 is the RPY 1973, in which Small originally put forwarded the method of co-citation analysis as a measurement of the correlation between two scientific papers (Small 1973). Co-citation analysis, as a milestone in iMetrics research, have been widespread and successfully utilized for detecting the research fronts and hot spots of a domain or topic in natural and social sciences.

After eight years, the second peak occurred referring to the variant of co-citation that is also co-citation analysis of scientific papers but on the author levels. White and Belver (1973) named it as author co-citation and applied it to measure the intellectual structure of a scientific domain or topic. The next significant peak happened in 1983 and it refer to a publication published by (Callon et.al. 1983). The contribution of their work is the introduction of co-word analysis, which can be treated as another variant of co-citation analysis conducted at the keyword level. However, the co-occurrence relationship between keywords were also considered except for co-citation relationship in the co-word analysis.

The fourth peak happened in 1988 is especially based on the book 《文献计量学》 ("Bibliometrics") published by the Science and Technology Literature Publishing House, Beijing China (Qiu 1988). The author of this book is Qiu J, who have been famous for his outstanding contribution to the development of iMetrics in China. In his book, Qiu J systematically introduced the basic theories and methodologies of the field of bibliometrics to Chinese readers. Despite most part of this book is the Chinese translation of research works from abroad, it is undeniable that the book has an irreplaceable positive effective on the origin and development of iMetrics in China. The book "Bibliometrics" has been reprinted in 1983, 1985, 1988 and has been used by more than 10 key universities in China as their teaching materials for the undergraduates and graduates.

The next peak in 1997 dates back to a paper "Informetric analyses on the world wide web: methodological approaches to 'webometrics'" by (Almind and Ingwersen 1997), where the definition and methods of webometrics were proposed. Intrinsically, webometrics was a variant of scientometrics or bibliometrics in the new era of Internet.

Finally, the last peak in the period is the RPY 2000, referring to a series of articles written in Chinese by Qiu J. In these papers, Qiu introduced the definition, development and evolution of the field of informetrics (Qiu 2000). This series of articles in a Chinese journal has been integrated into a book named 《信息计量学》 ("Informetrics") and published by the Wuhan University Press in 2007 (Qiu 2007).

*Stage three (2001-2017): The high-speed development of iMetrics in China*

In 21st century, the iMetrics in China began to develop with expansion, the number of publications, cited references, and citations all swift increased (Figure 3 and Figure 8). The total number of cited references during the period of 2001-2017 is 54,362 after deduplication. Although the citation times of references has not been steady, three obvious peaks in the Figure 8 have happened respectively in 2006, 2008 and 2013, which were significant to the development of the iMetrics in China. Table 4 shows the detailed information about the seminal works in the peaks during 2001-2017. Note that we use cited times of cited references instead of the percentage, since large number of references' citations during this period are only one time, which makes the value of percentage very low.

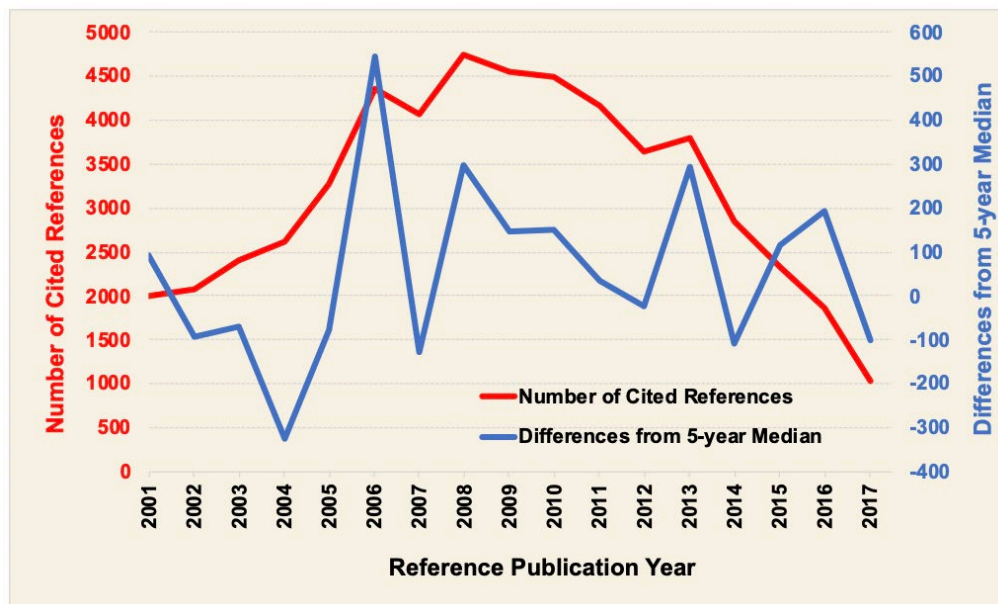

Figure 8. The reference publication year spectroscopy of iMetrics in China (2001-2017). Notes: the seminal works related to iMetrics in China were published in 2006, 2008 and 2013.

Table 4. Details about the significant peaks for iMetrics in China between 2000-2017.

| NO | RPY | Most Cited Reference | Cited times | Document Type |
|---|---|---|---|---|
| 1 | 2006 | Chen, C. (2006). CiteSpace II: Detecting and visualizing emerging trends and transient patterns in scientific literature. Journal of the American Society for information Science and Technology, 57(3), 359-377. | 196 | Journal |
| 2 | 2006 | 冯璐, 冷伏海. (2006). 共词分析方法理论进展. 中国图书馆学报, 32(3), 88-92. *[Feng, L., & Leng, F. (2006). The theoretical progress of co-word analysis. Journal of Library Science in China, 32(3), 88-92.]* | 152 | Journal |
| 3 | 2008 | 苏新宁. (2008). 构建人文社会科学学术期刊评价体系. 东岳论丛, (1), 35-42. *[Su X. N. (2008). Constructing an evaluation system for academic journals in humanities and social sciences. Dongyue Tribune, (1), 35-42.]* | 111 | Journal |
| 4 | 2013 | 邱均平, 余厚强. (2013). 替代计量学的提出过程和研究进展. 图书情报工作, 57(19), 5-12. *[Qiu J, & Yu H. (2013). The putting forward process and research progress of Altmetrics. Library and Information Service, 57(19), 5-12.]* | 79 | Journal |

The first peak happened in 2006 refers to two articles. The first article was "CiteSpace II: Detecting and visualizing emerging trends and transient patterns in scientific literature" by (Chen 2006), which was published in Journal of the American Society for information Science and Technology and cited by 196 times, only accounting 4.5% of the total citations of references published in 2006. CiteSpace, the most popular tool for knowledge mapping employed in China, was introduced in Chen's paper. The second article was a Chinese article entitled "The theoretical progress of co-word analysis" in the Journal of library science in China (Feng and Leng 2006). In this article, the authors systematically introduced the method of co-word analysis from three different perspectives, including the inclusion index and proximity index, the strategic diagram, and the database tomography (Feng and Leng 2006). This article has been the most cited article on co-word analysis in China.

The second peak happened in RPY 2008, referring to Su's "Constructing an evaluation system for academic journals in humanities and social sciences" (translation from Chinese), in which the construction of the China Social Sciences Citation Index (CSSCI) was introduced (Su 2008). The CSSCI has become one of the most authority academic database for humanities and social sciences in China and is also a significant data source for iMetrics research in China (Su et al. 2014).

Eventually, the last peak in this period is the RPY 2013, referring to a Chinese article entitled "The putting forward progress and research progress of Altmetrics" by (Qiu and Yu 2013). This article was the first article on Altmetrics published in Chinese, which introduced the historical roots and research progress of Altmetrics in the world. It has also become the most cited article on Altmetrics in China.

**Discussion and conclusion**

This study explored how to investigate the origin and evolution of research fields in China, using iMetrics in China as a case study. It is of paramount significance for identifying scientific paradigms shifts during the historical evolutionary path of a Chinese research field. The main contribution of this paper is two folds. First, this paper introduced RootCite (a Python-based tool for RPYS analysis on Chinese publications), showcased and verified how it can aid the task of locating seminal works in the historical evolutionary path of a Chinese field. Second, this paper examined the historical roots and seminal works of iMetrics in China using RootCite, which could be valuable for extending RPYS for countries with other languages. A total of 16 significant peaks referring to 16 seminal works (13 in English and 3 in Chinese) were identified during 1900-2017, which is characterized by three stages: budding (before 1900), formation (1971-2000), and development and expansion (2001-2017).

The results demonstrated that RootCite can be successfully used for identifying the origin and evolution of a given field in China with the supervision of domain experts. The research findings on iMetrics in China can be summarized as follows: Before 1970, iMetrics in China was in its budding stage and lacked the original and systematic research achievements. The earliest seminal work was written in English and entitled "The history of comparative anatomy: Part I.—A statistical analysis of the literature", which was considered as the start point of the field of bibliometrics (Cole and Eales 1917). Then, the classical works on theories and methods of bibliometric and scientometrics, such as Lotka's law (1926), Braford's law (1934), literature obsolescence (1944), citation index (1955) and bibliographic coupling (1963), were successively translated, introduced and absorbed by Chinese scientists.

In 1971-2000, iMetrics in China stepped into its formative stage, the terms "informetrics" and "webometrics" were formally defined (Milojević and Leydesdorff 2013) and introduced into China (Qiu et al. 2003). Approaches including co-citation analysis based on science citation index (SCI) and co-word analysis had been widely used by the iMetricians in China. It is worth noting that two Chinese works written by Qiu had made a huge positive contribution to the dissemination and development of iMetrics in China. These two works are 《文献计量学》 ("Bibliometrics") in 1988 (Qiu 1988) and 《信息计量学》 ("Informetrics") in 2000 (Qiu 2000), marking that bibliometrics and informetrics became the formal courses for the LIS students in China. In this stage, the number of publications of iMetrics in China showed a steadily growth. These Chinese articles not only introduced and reviewed the researches abroad, but also included the studies that applied iMetrics approaches to the Chinese materials or the exploration of the applicability of iMetrics theories on Chinese literatures (Qiu et al. 2003; Yang et al. 2019).

In the last stage (2001-2017), iMetrics in China had undergone a high-speed development and entered into its maturity. A scientific mapping software (i.e., CiteSpace) was introduced by (Chen 2006). CiteSpace can identify research trends and fronts of a given field; as its powerful visualization ability and the acceptance of Chinese bibliographic information, it becomes one of the most popular bibliometric software in China. Meanwhile, the China Social Science Index (CSSCI) was designed and developed in this stage, which provide a significant data source for iMetrics research in China (Su et al. 2014). Besides, the research progress of the methods of co-word analysis and Altmetrics in the world had been introduced into China (Feng and Leng 2006; Qiu and Yu 2013).

Compared to the findings of the research of Leydesdorff et al. (2014), in which the historical roots were respectively investigated using four datasets, that is, Scientometrics, Journal of Informetrics, JASIST-I (a subset of JASIST on iMetrics), and all three journals, as shown in Table 5, we can make two observations. First, the most significant peak of iMetrics in China happened in 1965 with a paper entitled "Networks of scientific papers" written by Price (1965) in English, different from the English world (Price's "Little science, big science"). This indicates that the iMetrics in China may have focused more on the specific theories and methods on the article level, while iMetrics in the English world might have paid more attention on the science of science. Second, all the three most significant peaks of iMetrics in China are in the set of that of iMetrics in the English world, indicating iMetrics in China rooted in the same contributions as the English world but it has its own characteristics. The pioneers of iMetrics in China have paid more attetion on applied aspect (e.g, paper networks and citation analysis), while the English world have deeped into the basic theory of this field (e.g., Lotka's law and Braford's law). Furthermore, despite most peaks in the historical evolutionary path of iMetrics in China happened abroad, there were still several Chinese works (e.g. Qiu's "Bibliometrics" and Su's CSSCI) that have an irreplaceable and positive effects on the development and evolution of iMetrics in China, whose contribution for the field should not be ignored. In a word, it is necessary to take Chinese dataset into consideration when identifying the seminal works in the historical evolutionary path of research fields in China.

**Limitations and future work**

There are several limitations in this paper. First, the publications relating to iMetrics in China are only from 1998 in the CSSCI database. However, as the RPYS method is rather robust approach that locates the seminal works from the perspective of cited references, the main results should not be affected. Moreover, the majority of the publications for iMetrics in China appeared after 2000, and the time period of cited references is from 1882 to 2018. Second, RootCite is still in its beta-version. Like the RPYS.exe (Marx and Bornmann 2014), it only provides users with the standard RPYS analysis and has no its own visualization module. In the future work, we will further optimize the RootCite and provide users with more useful features to solve their bibliometric tasks, such as multi-RPYS analysis and "sleeping beauty" recognition. Besides, in the future, we also plan to conduct the RPYS analysis on other research fields in China using RootCite to identify the seminal works in their historical evolutionary paths.

Table 5. The three most significant peaks of iMetrics in different raw data during the period of 1900-1970.

| Raw Data | First peak | Second peak | Third Peak |
|---|---|---|---|
| CSSCI | Price, D. J. de Solla. (1965). Networks of scientific papers. *Science, 149*(3683), 510-515. | Garfield, E. (1955). Citation indexes for science: A new dimension in documentation through association of ideas. *Science, 122*(3159), 108-111. | Kessler M.M. (1963). Bibliographic coupling between scientific papers. *American Documentation, 14(1),* 10-25. |
| Scientometrics | (1) Price, D. J. de Solla. (1963). Little science, big science. New York: Columbia University Press. (2) Kessler M.M. (1963). Bibliographic coupling between scientific papers. *American Documentation, 14(1),* 10-25. (3) Garfield E. (1963). Citation Indexes in sociological and historical research. *American Documentation, 14(4),* 289-291. | Lotka, A.J. (1926). The frequency distribution of scientific productivity. *Journal of the Washington Academic of Science, 16(12),* 317-323. | (1) Merton, R. K. (1957). Priorities in scientific discovery: A chapter in sociology of science. *American Sociology Review*, *22(6),* 635-639. (2) Farrell, M. J. (1957). The measurement of productive efficiency. *Journal of the Royal Statistical Society. Series A (General)*, *120*(3), 253-290. |
| Journal of Informetrics | (1) Price, D. J. de Solla. (1963). Little science, big science. New York: Columbia University Press. (2) Kessler M.M. (1963). Bibliographic coupling between scientific papers. *American Documentation, 14(1),* 10-25. (3) Kessler M.M. (1963). Bibliographic coupling extended in time: Ten case histories. *Information Storage & Retrieval, 1*(4), 169-187. | Garfield, E. (1955). Citation indexes for science: A new dimension in documentation through association of ideas. *Science, 122(3159),* 108-111. | Merton, R. K. (1968). The Matthew effect in science: The reward and communication systems of science are considered. *Science, 159*(3810), 56-63. |
| JASIST - I | (1) Price, D. J. de Solla. (1963). Little science, big science. New York: Columbia University Press.<br>(2) Kessler M.M. (1963). Bibliographic coupling between scientific papers. *American Documentation, 14(1),* 10-25.<br>(3) Kessler M.M. (1963). Bibliographic coupling extended in time: Ten case histories. *Information Storage & Retrieval, 1*(4), 169-187. | (1) Price, D. J. de Solla. (1965). Networks of scientific papers. Science, 149(3683), 510-515.<br>(2) Kaplan, N. (1965). The norms of citation behavior: Prolegomena to the footnote. *American Documentation, 16*(3), 179-184. | Lotka, A.J. (1926). The frequency distribution of scientific productivity. *Journal of the Washington Academic of Science, 16(12),* 317-323. |
| Three Journals | (1) Price, D. J. de Solla. (1963). Little science, big science. New York: Columbia University Press.<br>(2) Kessler M.M. (1963). Bibliographic coupling between scientific papers. *American Documentation, 14(1),* 10-25.<br>(3) Garfield, E, & Sher, I. (1963). New factors in the evaluation of scientific literature through citation indexing. *American Documentation, 14(3),* 195-201.<br>(4) Garfield E. (1963). Citation Indexes in sociological and historical research. *American Documentation, 14(4),* 289-291. | Lotka, A.J. (1926). The frequency distribution of scientific productivity. *Journal of the Washington Academic of Science, 16(12),* 317-323. | Bradford, S.C. (1934). Sources of information on specific subjects. *Engineering, 137,* 85-86. |

## Acknowledgements

The financial support provided by Wuhan University (student exchange program) during the visit by Xuli Tang to the Indiana University Bloomington is acknowledged. The financial support provided by the China Scholarship Council (CSC) during a visit of Xin Li to The University of Texas at Austin (No. 201806270047) is acknowledged. The authors are also grateful to the anonymous referees and editors for their invaluable and insightful comments.

## Supplementary Information

### S1. Search strategy of this study

*All Fields = ('bibliometrics' OR 'webometrics' OR 'informetrics' OR 'scientometrics' OR 'knowledge metrics' OR 'citation analysis' OR 'altmetrics' OR 'co-word analysis' OR 'journal evaluation' OR 'paper evaluation' OR 'scientific evaluation' OR 'academic impact' OR 'h index' OR 'university rank' OR 'open access') AND publication year = (1998-2017) AND article type = ('article' or 'review').*

所有字段=（'文献计量' OR '网络计量' OR '信息计量' OR '科学计量' OR '知识计量' OR '引文分析' OR '补充计量学' OR '替代计量学' OR '共词分析' OR '期刊评价' OR '论文评价' OR '科研评价' OR '学术影响力' OR 'H 指数' OR '大学排名' OR '开放获取'） AND 出版年份＝（1998-2017）AND 文献类型＝（'论文' or '综述'）

### S2. An example of the bibliographic information of an article downloaded from the CSSCI

[Figure: screenshot of 1998-1.txt showing fields: Chinese title 【来源篇名】关于文献老化研究中若干问题的思考; English title 【英文篇名】Thinking about Some Problems in the Research on Document Obsolescence; Authors 【来源作者】俞培果; Funding 【基 金】; Journal 【期 刊】情报理论与实践; First affiliation 【第一机构】西南工学院; Affiliations 【机构名称】[俞培果]西南工学院.; First author 【第一作者】俞培果; Classification code 【中图类号】G350; Publication year, issues 【年代卷期】1998,21(030):144-146,131; Keywords 【关 键 词】文献老化/文献计量学/文献学; Level of funding 【基金类别】; Cited references 【参考文献】 1. Brookes,B.C..The Growth,Utility,and Obsolescence of Scientific Periodical Literature.Journal of Documentation.1970.26(4):283-294; 2. Burton,R.E..The Half Life of Some Scientific and Technical Literatures.American Documentation.1960.11(1); 3. Cosnell,C.F..Rate of Obsolescence in College Library Book Collections by an Analysis of Thrbr,New York Univ,1934; 4. De S Price D..Nasurari de Referinte Bibliografice(Citate)Sn Domeniul Stiintelor, dens Structur,.Studii Sicercetari de Documentare.1970.12(3); 5. Line,M.B..Obsolescence and Changes in the Use of Literature With Time.Journal of Documentation.1974.30(4):283-350; 6. 丁学东.文献计量学基础.北京:北京大学出版社, 1993; 7. 李莲馥.科技文献老化的概念定义.国外情报科学.1985. (1); 8. 徐新民.科学交流与情报学.北京:科学技术文献出版社, 1980; 9. 严怡民.情报学概论.武汉:武汉大学出版社, 1983; 10. 俞培果.馆藏年代分布对藏书老化测定的影响.情报刊.1991.12 (1); 11. 俞培果.科技文献老化的历时研究.情报业务研究.1993.10 (2); 12. 俞培果.两种文献老化测度方法的比较研究.情报业务研究.1993.10 (4); 13. 俞培果.文献产生的年代分对引文年代分布的影响.情报业务研究.1991.8 (3); 14. 俞培果.引文率指标的测度意义和测度性质研究.情报业务研究.1992.9 (2); 15. 中国人民大学哲学系逻辑教研室.形式逻辑.北京:中国人民大学出版社, 1980]

### S3. How to use RootCite

AS we can see from Figure 1, RootCite contains four modules listed in the right of its interface, that is, (1) file module ("creat"); (2) preprocessing module ("readCSSCI and readWOS"); (3) rpys module ("rpys and year") and (4) deduplication module ("deduplication"). To investigate the historical roots and evolution of iMetrics research in China, the following four detached steps have been adopted with RootCite, as shown in Figure 2.

Step 1: **generating a new project**. Double click the RootCite to start it up and click the create button in the file module to create a new project, then you can find a folder called RootCiteProject including two subfolders (data_cssci and data_wos) in current directory.

Step 2: **preprocessing**. Put one or more plain texts downloaded from the CSSI to the data_cssci folder and click the readCSSCI in the preprocessing module to extract all cited references.

Step 3: **computing the value of rpys and median**. Click the deduplication button to deduplicate the variants and misspelling of cited references, and click the rpys button, then rpys_cssci.csv and median_cssci.csv will be generated. Thereafter, click the year button, and the file result_cssci.csv will be generated.

Step 4: **visualization and analysis**. Using Excel to open median_cssci.csv, we can draw the reference publication year spectroscopy that can be seen from Figure 4 and identify peak RPYs; then we can find the details about the significant publications in a specific peak year with result_cssci.csv.